\documentclass[conference,unknownkeysallowed]{IEEEtran}

\IEEEoverridecommandlockouts
\def\BibTeX{{\rm B\kern-.05em{\sc i\kern-.025em b}\kern-.08emT\kern-.1667em\lower.7ex\hbox{E}\kern-.125emX}}

\usepackage[font={small, it}]{caption}    
\usepackage{todonotes}    
\usepackage{nicefrac}
\usepackage{siunitx}
\usepackage{array,framed}
\usepackage{booktabs}
\usepackage{fancyvrb}
\usepackage{graphicx}

\usepackage{
  color,
  float,
  epsfig,
  wrapfig,
  graphics,
  graphicx,
  subcaption
}
\usepackage{textcomp,amssymb}
\usepackage{setspace}
\usepackage{latexsym,fancyhdr,url}
\usepackage{enumerate}
\usepackage{algorithm2e}
\usepackage{algpseudocode}
\usepackage{xparse} 
\usepackage{xspace}
\usepackage{multirow}
\usepackage{csvsimple}
\usepackage{balance}
\usepackage{graphicx}
\usepackage{subcaption}

\usepackage{
  tikz,
  pgfplots,
  pgfplotstable
}
\usepackage{hyperref}

\usetikzlibrary{
  shapes.geometric,
  arrows,
  external,
  pgfplots.groupplots,
  matrix
}

\pgfplotsset{compat=1.9}

\usepackage{mathtools,}

\DeclareMathAlphabet{\mathcal}{OMS}{cmsy}{m}{n}

\DeclareGraphicsExtensions{%
    .png,.PNG,%
    .pdf,.PDF,%
    .jpg,.mps,.jpeg,.jbig2,.jb2,.JPG,.JPEG,.JBIG2,.JB2}

\usepackage{xparse}
\newcommand{\bnm}{\begin{newmath}}
\newcommand{\enm}{\end{newmath}}

\newcommand{\bea}{\begin{eqnarray*}}%
\newcommand{\eea}{\end{eqnarray*}}%

\newcommand{\bne}{\begin{newequation}}
\newcommand{\ene}{\end{newequation}}

\newcommand{\bal}{\begin{newalign}}
\newcommand{\eal}{\end{newalign}}

\newenvironment{newalign}{\begin{align}%
\setlength{\abovedisplayskip}{4pt}%
\setlength{\belowdisplayskip}{4pt}%
\setlength{\abovedisplayshortskip}{6pt}%
\setlength{\belowdisplayshortskip}{6pt} }{\end{align}}

\newenvironment{newmath}{\begin{displaymath}%
\setlength{\abovedisplayskip}{4pt}%
\setlength{\belowdisplayskip}{4pt}%
\setlength{\abovedisplayshortskip}{6pt}%
\setlength{\belowdisplayshortskip}{6pt} }{\end{displaymath}}

\newenvironment{newequation}{\begin{equation}%
\setlength{\abovedisplayskip}{4pt}%
\setlength{\belowdisplayskip}{4pt}%
\setlength{\abovedisplayshortskip}{6pt}%
\setlength{\belowdisplayshortskip}{6pt} }{\end{equation}}

\newcounter{ctr}

\newcounter{mytable}
\def\mytable{\begin{centering}\refstepcounter{mytable}}
\def\endmytable{\end{centering}}

\newcounter{myfig}
\def\myfig{\begin{centering}\refstepcounter{myfig}}
\def\endmyfig{\end{centering}}

\newlength{\saveparindent}
\newlength{\saveparskip}
\newcommand{\E}{{\rm I\kern-.3em E}}

\renewcommand{\eqref}[1]{\mbox{Equation~(\ref{#1})}}

\def \part {part}

\renewcommand{\paragraph}[1]{\vspace*{6pt}\noindent\textbf{#1}\;}

\def \blackslug{\hbox{\hskip 1pt \vrule width 4pt height 8pt
    depth 1.5pt \hskip 1pt}}
\def \qed{\quad\blackslug\lower 8.5pt\null\par}

\newcounter{mynote}[section]

\newcommand\ignore[1]{}

\newcounter{rcnote}[section]

\newcounter{mrnote}[section]

\newcounter{fknote}[section]

\newcounter{anote}[section]

\DeclareMathSymbol{\mlq}{\mathord}{operators}{``}
\DeclareMathSymbol{\mrq}{\mathord}{operators}{`'}

\newcommand{\rhf}[2]{R_{f, \gamma}}

\DeclareDocumentCommand{\edist}{o o}{
  \ensuremath{
    \IfNoValueTF{#1}{{d}}{{\sf d}(#1,#2)}
  }
}

\newcommand{\olrk}[1]{\ifx\nursymbol#1\else\!\!\mskip4.5mu plus 0.5mu\left(\mskip0.5mu plus0.5mu #1\mskip1.5mu plus0.5mu \right)\fi}

\NewDocumentCommand{\indseq}{ O{1} O{r} }{{#1}\ldots {#2}}

\usepackage[absolute]{textpos}

\makeatletter
\def\ps@IEEEtitlepagestyle{%
  \def\@oddfoot{\mycopyrightnotice}%
  \def\@evenfoot{}%
}
\def\mycopyrightnotice{%
  {\footnotesize  This paper has been accepted for publication by the 18th IEEE Annual Consumer Communications \& Networking Conference (CCNC). The copyright is with the IEEE. \hfill}
  \gdef\mycopyrightnotice{}
}

\begin{document}


\fancyhead{}
\def\thetitle{Irregular Metronomes as Assistive Devices to Promote Healthy Gait Patterns}
\title{\thetitle}

\author{
\IEEEauthorblockN{Aaron D. Likens\IEEEauthorrefmark{1},
Spyridon Mastorakis\IEEEauthorrefmark{2},
Andreas Skiadopoulos\IEEEauthorrefmark{1}, \\
Jenny A. Kent\IEEEauthorrefmark{3},
Md Washik Al Azad\IEEEauthorrefmark{2},
Nick Stergiou\IEEEauthorrefmark{1}
}
\IEEEauthorblockA{\IEEEauthorrefmark{1}Department of Biomechanics, University of Nebraska at Omaha, Omaha, USA\\
Email: alikens@unomaha.edu, askiadopoulos1@unomaha.edu, nstergiou@unomaha.edu}
\IEEEauthorblockA{\IEEEauthorrefmark{2}Computer Science Department, University of Nebraska at Omaha, Omaha, USA\\
Email: smastorakis@unomaha.edu, malazad@unomaha.edu}
\IEEEauthorblockA{\IEEEauthorrefmark{3}Physical Medicine and Rehabilitation, Northwestern University, Chicago, USA\\
Email: jenny.kent@northwestern.edu}
}



\maketitle
\IEEEpubidadjcol
\begin{abstract}

Older adults and people suffering from neurodegenerative disease often experience difficulty controlling gait during locomotion, ultimately increasing their risk of falling. To combat these effects, researchers and clinicians have used metronomes as assistive devices to improve movement timing in hopes of reducing their risk of falling. Historically, researchers in this area have relied on metronomes with isochronous interbeat intervals, which may be problematic because normal healthy gait varies considerably from one step to the next. More recently, researchers have advocated the use of irregular metronomes embedded with statistical properties found in healthy populations. In this paper, we explore the effect of both regular and irregular metronomes on many statistical properties of interstride intervals. Furthermore, we investigate how these properties react to mechanical perturbation in the form of a halted treadmill belt while walking. Our results demonstrate that metronomes that are either isochronous or random metronome break down the inherent structure of healthy gait. Metronomes with statistical properties similar to healthy gait seem to preserve those properties, despite a strong mechanical perturbation. We discuss the future development of this work in the context of networked augmented reality metronome devices. 

\end{abstract}


\section{Introduction} 
\label{sec:intro}

A number of clinical populations experience difficulty in controlling gait during locomotion, and such difficulties increase the risk of falling. Hence, there is a strong interest in developing easy-to-use technology to improve gait, and ultimately, prevent falls. In recent years, a number of researchers and clinicians have noted the benefit of using metronomes to improve the control of human gait in cases of aging and neurodegenerative disease such as Parkinson's Disease~\cite{cubo2004short,vaz2020gait}. Similar improvements have been observed in athletic training, where the effect of metronomic training has been suggested to exhibit a general effect on motor coordination.  Hence, the use of metronomes shows great promise in these clinical and exercise science settings~\cite{kim2019effects}. An issue, though, is that those studies have largely used isochronous interbeat intervals for training and rehabilitation. An alternative approach is the use of irregular metronomes based on the statistical regularities observed in gait patterns of young, healthy people~\cite{hunt2014influence,kaipust2013gait}. Results from our research team suggest that this alternative approach may be effective in restoring healthy gait patterns in older adults~\cite{vaz2020gait,kaipust2013gait}. 

The contribution of this paper is two-fold. First, we present a laboratory study that employs irregular metronomes within the context of mechanical perturbations to gait. Second, we discuss our ongoing work re-thinking these metronomes as devices with advanced networking capabilities, enhanced with immersive technologies, such as Augmented Reality. Our paper is organized as follows. In Section~\ref{sec:relwork}, we give a brief background and discuss related work, while in Section~\ref{sec:experiments}, we present our methodology. In Section~\ref{sec:results}, we present our evaluation and, in Section~\ref{sec:nextgen}, we discuss directions on developing the next-generation of network-capable metronomes. Finally, in Section~\ref{sec:conclusion}, we conclude our work.

\section{Background and Related Work}
\label{sec:relwork}

Variability is a fundamental feature of both human movements and the underlying physiological processes that support them~\cite{Aks2002Memory, Stergiou2011Human, Peng1995Fractal, Peng1995Quantification, Ivanov2001From, Kello2010Scaling, cavanaugh2017multifractality}. For example, gait variability reflects how gait parameters (e.g., stride time, stride length) change over many successive steps. Physiological variability may be observed by successive observations of physiological events such as the timing of breaths or heart beats. The study of variability in overt movements and physiological events has revealed systematic patterns of variability associated with both health and disease. Historically, variability was thought to be synonymous with error such as a faulty execution of an intended motor program~\cite{Fitts1954information,Reason1990Human,Schmidt2003Motor}. Evidence supporting this point of view is the frequent observation that novices, older adults, and those with neurodegenerative diseases often produce greater variability than young, healthy, experts. From this point of view, variability is something to be reduced or eliminated. This perspective has become increasingly difficult to support given the growing evidence that structured variability is essential for a healthy, adaptive system~\cite{Stergiou2006Optimal,harrison2015complex}. Elimination of variability introduces the possibility of a system that is inflexible in the face of novel circumstances. Novel circumstances are the rule in locomotion which involves an ever-changing environment. In contrast, optimal variability should strike a balance between complexity (i.e., behavioral richness) and predictability. Too much predictability leads to robotic movements; too little predictability leads to randomness. Both overly predictable and overly random behavior has been associated loss of adaptability associated with aging and disease.
\subsection{Fractality in Human Movements}
Human movements entail the coordination of many nested neuromuscular processes~\cite{cavanaugh2017multifractality,kelty2013tutorial,Likens2014Neural,Likens2015Experimental,Goldberger2002Fractal, Ivanov2001From,harrison2015complex}. The successful interaction of the many components that make up such a system are reflected in the time varying properties (i.e., variability) observed in movement. Movement and physiological patterns represent a form of variability known as "multifractality" that has been suggested to represent the coordination of motor degrees of freedom~\cite{cavanaugh2017multifractality}. Fractals (aka monofractals) are geometric objects that are self-similar over different scales of analysis~\cite{Goldberger2002Fractal}. Fractals are observable from time series data (e.g., a sequence of stride times) by investigating how statistical properties (e.g., variance) change according to scale. When the logarithm of variance increases linearly with the logarithm of scale, the time series is said to exhibit fractal scaling. The slope that relates these quantities is the fractal scaling exponent, sometimes referred to as the Hurst exponent, $H$. Monofractal scaling suggests that single scaling exponent is sufficient for characterizing the dynamics of gait events. Furthermore, monofractal scaling does not in itself support the conclusion of interactions among the time scales that make up movement. Hence, this one-size fits all approach may not reflect the adaptability needed in human motor control. Multifractality, on the other hand, suggests that fractal patterns may be context dependent, changing to reflect the adaptive nature of the human neuromuscular system. 

Although there are exceptions~\cite{ashkenazy2002stochastic,dutta2013multifractal,west2005multifractal}, the majority of work investigating fractal characteristics in human gait have been conducted from a monofractal perspective~\cite{Hausdorff1995Is,hausdorff1996fractal,hunt2014influence,kaipust2013gait}. Furthermore, although the influence of metronomes on monofractal properties of gait are well known~\cite{kaipust2013gait,hunt2014influence,hausdorff1996fractal}, we are not aware of any studies that have systematically investigated the effect of metronomes on mulitfractal patterns in observed in human gait. The current work addresses this gap in the literature while also providing critical information regarding the role of gait fractality in resilience to mechanical perturbations. 
\subsection{Metronomes as Assistive Devices to Prevent Falls}
Auditory and visual metronomes have emerged as promising rehabilitation tools for restoring healthy motor patterns in a number of domains including stroke, Parkinson’s disease, aging~\cite{hollands2015feasibility,vaz2020gait,hunt2014influence,baker2008effect,spaulding2013cueing}. For example, pacing people with an isochronous metronomes seems to increase gait speed and stride length~\cite{spaulding2013cueing}. At issue, though, is that studies of this nature often employ isochronous metronomes, which runs contrary to known properties of human movement variability~\cite{kaipust2013gait,hunt2014influence, hausdorff1996fractal}. This is problematic because, while some studies show improvement in gait parameters using isochronous metronomes, other studies have shown that isochronous metronomes may actually alter healthy dynamics in such a way that mimics pathological gait variability~\cite{kaipust2013gait,hunt2014influence,vaz2020gait}. In contrast, using metronomes that exhibit the same properties as found in healthy gait does not degrade healthy gait dynamics~\cite{hunt2014influence} and may have exert a restorative effect on gait patterns in older adults~\cite{van2016efficacy}.   

\subsection{The current study}
In this paper, we present a novel study in which we demonstrate that metronomes may be effective in reducing the risk of falling due to an unexpected perturbation. Participants walked on a treadmill while either being paced by one of several metronome types or walking at a self-selected comfortable pace. During the trial, participants experienced a mechanical perturbation (a treadmill belt was briefly halted). We hypothesized that a metronome with statistical characteristics similar to those of healthy, self-paced walking would exhibit more resilience to perturbation as compared with other metronome types. 

\section{Methodology}
\label{sec:experiments}

\subsection{Participants}
Thirty-three healthy young adults (age: 19 – 30 years; height: 1.75 ± 0.91; weight: 71 ± 0.1 kg; 13 females, 20 males) volunteered to participate in this study. All participants reported normal or corrected-to-normal vision and hearing, and the ability to walk for $>=$ 45 consecutive minutes. Participants reported no neurological, movement or vestibular disorder, dizziness, impairments to vision or hearing, current musculoskeletal injury or pain. Participants had no cardiovascular events, and had not experienced surgery within the previous 6 months, and were not pregnant. The institutional review board approved all procedures.

\subsection{Experimental protocol}
Walking trials took place on an instrumented, split belt treadmill (Bertec Corp., OH, USA). Participants were randomly assigned to one of four groups, corresponding to each of the four metronome conditions; no metronome (or none), isochronous, random, and 1/f noise. Participants wore tight-fitting clothing and were fitted with a ceiling-mounted harness. We determined self-selected walking speed by  increasing and decreasing treadmill speed until the participant reported that the speed was ‘faster than comfortable’ or ‘slower than comfortable’. We averaged the first three ‘faster’ and ‘slower’ to estimate a comfortable, self-selected speed and used that value for the duration of the session.

\subsection{Baseline trial}
Participants walked on the treadmill for 25 minutes at the self-selected walking speed but without a metronome. After the baseline trial, participants rested for 20 minutes. Right foot contact events from the last 3 minutes of the baseline trial were used to compute inter-stride intervals (ISIs; mean and SD) based on the trajectory of the vertical velocity of the left and right heel markers.

\subsection{Stimuli creation}
Baseline gait characteristics for each participant were used to construct 45 minute metronomes, appropriate to group, using MATLAB (The MathWorks Inc, Natick, MA, USA). Inter-beat (IBI) intervals for the isochronous condition were set to each participants’ mean baseline ISI. For the random condition, a random sequence of values between -1 and 1 was generated normalized baseline ISIs means and variances. 1/f noise time series were generated using custom MATLAB scaled in a manner similar to the random condition. Metronome intervals were converted to 4-beat drum patterns wherein beats were sounded by a closed hi-hat. A base drum and snare were played on the first and third beats, respectively. The bass drum signaled start of stride; the snare drum signaled contralateral limb foot contact. Stimuli were converted to MIDI sequences, and played by drum generator app (Drum Studio, Rollerchimp, Sydney, Australia). Participants in the metronome condition groups were asked to time their steps to beat of the metronome while walking. Before beginning experimental trials, participants demonstrated synchronizing their steps with beats. If initially unable to synchronize, researchers coached them and then participants practiced 2 additional minutes. Some participants required minor coaching but were all able to walk in time with the beat.

\subsection{Perturbation trial}
Thirty-nine retroreflective markers were affixed to anatomical landmarks to define a nine-segment (left and right foot, left and right shank, left and right thigh, pelvis, trunk, head) mechanical model of the human body. Clusters of rigid reflective markers were also placed on the thigh and lower leg. Participants walked for 45-minute in one of three metronome conditions (1/f noise, random, or isochronous) or at a self-selected speed. Metronome beats were played speakers at the same volume for all conditions. Twenty-five minutes into the trial, we halted one treadmill belt for 500ms, providing a mechanical perturbation meant to approximate a trip. The perturbation was timed to when the ankle of the dominant limb in swing passed the ankle of the support limb. Treadmill acceleration and deceleration was set to 3 ms-2. Participants were asked to resume walking after the normal belt movement resumed. The trial continued for 20 more minutes. Kinematic data was recorded at 100 Hz using an 8-camera, 3D motion capture system (Vicon Motion Systems, Oxford, UK).  The timing and duration of the treadmill perturbation were controlled by D-Flow (Motek Medical BV, Amsterdam, The Netherlands). Participants only received a single perturbation to prevent learning effects. Participants were naive to the trip.

\subsection{Analysis}
Periods before (PRE) and after the perturbation (POST) were identified from the 45 minutes walking trials and the foot contact events were estimated based on the trajectory of the vertical velocity of the right heel marker using custom Matlab code. Foot contact events were also identified from the 25 minute baseline trial. ISIs were calculated as the time elapsed between subsequent heel strike events of the same foot. 

After extraction, ISIs were used as input to a time series method called Multifractal Detrended Fluctuation Analysis (MFDFA)~\cite{kantelhardt2002multifractal}. MFDFA is a generalization of well known Detrended Flucatuation Analysis \cite{peng1994mosaic} and has been used in numerous disciplines to study the correlation structure of time series data, including relevant areas such as physiology and movement science~\cite{Hausdorff1995Is,hausdorff1996fractal, hunt2014influence, kaipust2013gait,Goldberger2002Fractal, Likens2014Neural, Likens2015Experimental}. See~\cite{kelty2013tutorial,ihlen2012introduction} for tutorials on this and related methods. We used MFDFA to characterize the time-varying properties of gait at baseline and during each experimental task segment (PRE/POST perturbation). Kantelhardt et al.~\cite{kantelhardt2002multifractal} describe MFDFA as an algorithm involving five steps. 
\begin{itemize}
    \item Step 1 involves determining the \emph{profile} as the cumulative sum of a mean-detrended time series according to 
    \[Y(i)=\Sigma_{k=1}^i(x_k-\Bar{x}),  i = 1,...,N\]
    where $N$ is the length of the series, $x_k$, and $\Bar{x}$ is mean of $x_k$.
    \item Step 2 requires dividing $Y(i)$ into $int(N/s)$ non-overlapping segments of length $s$. Because $N$ will not always be a multiple of $s$, the procedure is repeated twice, once from each end of $Y(i)$ to avoid dropping data. This results in $2N_s$ segments.
    \item Step 3 calculates local trends in each of the $N_s$ segments by ordinary least squares regression. The variance is then estimated as:
    \[ F^2(v,s)=\frac{1}{s}\Sigma_{i=1}^{s}\{Y[(v-1)s+i]-y_v(i)\}^2 \]
    for each segment, $v, v = 1,...,N_s$ and
    \[ F^2(v,s)=\frac{1}{s}\Sigma_{i=1}^{s}\{Y[N-(v-N_s)s+i]-y_v(i)\}^2 \]
    for $v=N_s+1,...2N_s$. $y_v(i)$ is the local trend determined by OLS regression. Here, $y_v(i)$ is assumed to be a first order polynomial, but higher order polynomials may also be used\cite{kantelhardt2002multifractal}.
    \item Step 4 entails estimating the $q$th order fluctuation function by averaging over all $2N_s$ segments according to
    \[ F_q(s)=\{ \frac{1}{2N_s}\Sigma_{v=1}^{2N_s}[F^2(v,s)]^{\frac{q}{2}} \}^{\frac{1}{q}}\]
    where $q$ is a real number. When $q=2$, the procedure reduces to the standard DFA\cite{peng1994mosaic} procedure. Steps 2 through 4 are repeated for several values of $s$ where $s\geq m+2$, where $m$ is polynomial order of $y_v(i)$. In the current work, we used $s_{min}=16$ and $q=-10,-9,...10$.
    \item Step 5 identifies the scaling behavior of $F_q(s)$ by regressing its logarithm against the logarithm of $s$. If the time series exhibits long range correlation, then $F_q(s)$ increases like a power law as a function of $s$, $F_q(s) ~ s^h(q)$, where the generalized Hurst exponent, $h(q)$ is the slope of the log-log plot of $F_q(s)$ and $s$. In the case of of multifractality, the collection of slopes, $h(q)$ will not be equal for all $q$. When $s$ is large, only a few values are included in calculation of $F_q(s)$. Hence, by convention, we chose $s_{max}=N/4$ for the current analysis. Lastly, $h(q)$ as $q$ tends toward 0, so $h(0)$ is estimated according to:
    \[F_0(s)=\exp{\frac{1}{4N_s}\Sigma_{v=1}^{2N_s}\ln[F^2(v,s)]} = s^{h(0)}.\]
    Other scaling exponents may be calculated based on $h(q)$. See \cite{kantelhardt2002multifractal} for details.
\end{itemize}

After extracting $h(q)$ by applying MFDFA to each ISI time series, $h(q)$ were analyzed via linear mixed effect models (LMEs). Models were fit using forward selection of parameters starting with fixed effects, adding higher order interactions one at time. Model terms were retained based on significant likelihood ratio tests. 

\section{Evaluation}
\label{sec:results}
LME results are presented as an analysis of variance in Table~\ref{table:anova}. That table indicates that all main effects and interactions were significant. Main effects and two-way interactions are not interpreted further because their implication in a three-way interaction among time, group, and q-order. 
\begin{table}[ht]

\centering
\begin{tabular}{lrrrrrr}
  \hline
 & SS & MS & DF1 & DF2 & F \\ 
  \hline
q & 5.48 & 5.48 & 1.00 & 1879.60 & 268.53  \\ 
  g & 1.68 & 0.56 & 3.00 & 27.61 & 27.48  \\ 
  t & 22.81 & 11.40 & 2.00 & 1887.16 & 558.63 \\ 
  q$\times$g & 0.73 & 0.24 & 3.00 & 1879.60 & 11.87 \\ 
  q$\times$t & 0.22 & 0.11 & 2.00 & 1879.60 & 5.46 \\ 
  g$\times$t & 30.79 & 5.13 & 6.00 & 1886.43 & 251.36 \\ 
  q$\times$g$\times$t & 1.57 & 0.26 & 6.00 & 1879.60 & 12.83\\ 
   \hline
\end{tabular}
\caption{\label{table:anova}Type III Analysis of Variance Table adjusted with Kenward-Rogers method. Results suggest a three-way interaction among q-order (q), group (g), and time (t). All $p < .01$}
\vspace{+0.5cm}
\end{table}

\begin{figure}
\centering

\includegraphics[width=1\linewidth]{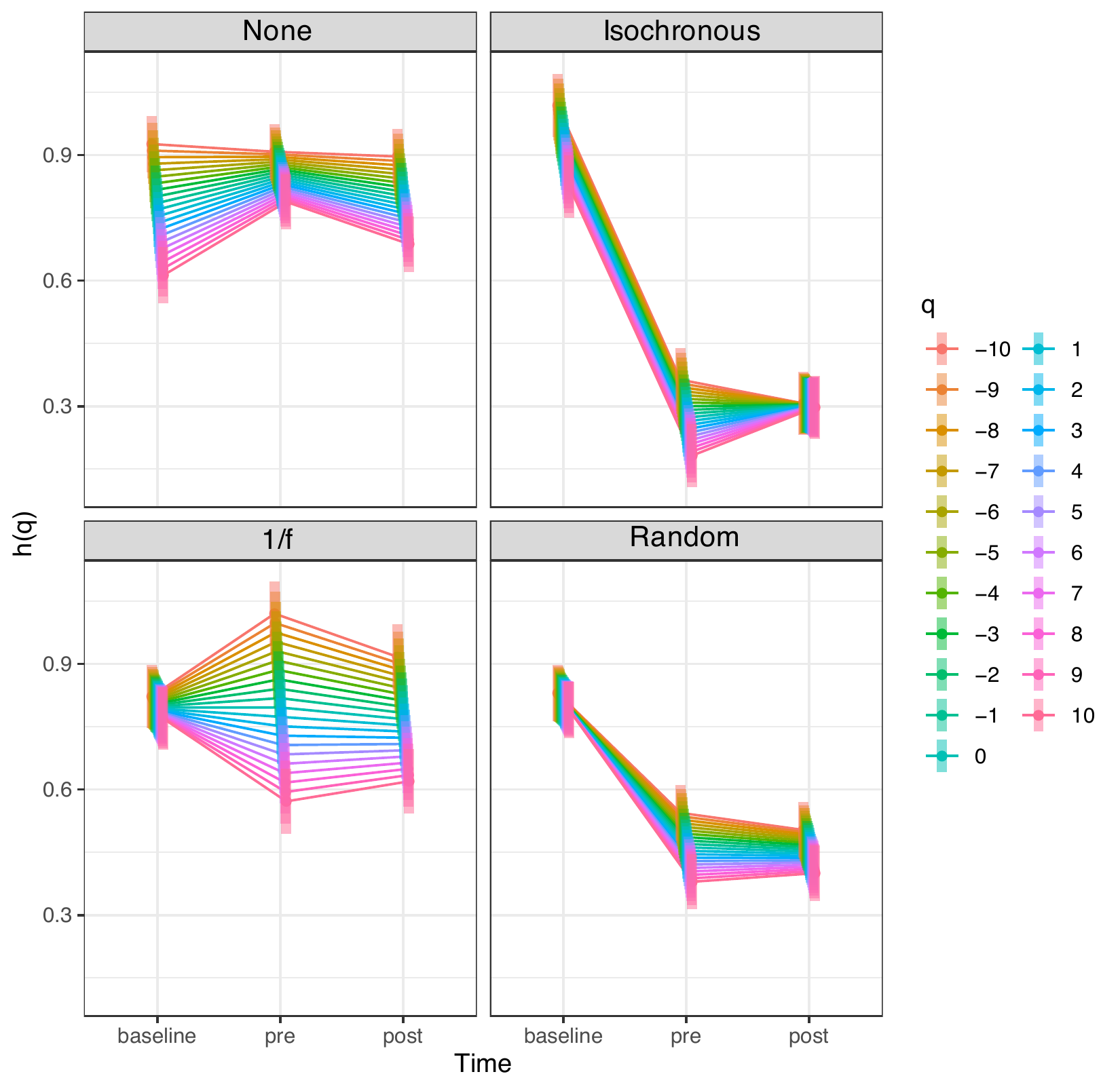}
\caption{Generalized Hurst exponent, $h(q)$, as a function of time and group.}
\label{fig:qorder} 

\vspace{+0.5cm}
\end{figure}

Model implied means are depicted in Figure~\ref{fig:qorder} as a way to visualize this interaction. Based on that figure, several general trends seem apparent. First, both the random and the isochronous metronomes reduced the central tendency of $h(q)$. On average, $h(q)$ appears lower in those conditions than either none or the 1/f conditions, replicating known trends in the literature\cite{hunt2014influence}. Second, the range of $h(q)$ appears to be larger when paced by a 1/f metronome than when paced by a either a random or isochronous metronome. To investigate the influence of group and time on $h(q)$, we performed a series of simple effects tests at extreme values of $q$ included in the MFDFA procedure i.e., (-10,10). These tests are presented in Table~\ref{table:hq_table}. When $q=-10$, small fluctuations dominate the estimation of fractal scaling. In that case, the none condition did not vary across time. However, large decreases in $h(q)$ were observed in the isochronous and random conditions, pre- and post-perturbation relative to the baseline. In contrast, the 1/f condition increased $h(q)$ relative to baseline. Only the 1/f condition showed a pre-post decline in $h(q)$, possibly due to floor effects in the other two conditions. When $q=10$, large fluctuations dominate the estimation of fractal scaling. In this case, all conditions showed differences in scaling behavior relative to baseline. However, only the none and isochronous conditions showed differences between pre- and post-perturbation. 

\begin{table}[ht]
\centering
\begin{tabular}{lrrrrl}
  \hline
contrast & estimate & SE & df & t & p \\ 
  \hline
\multicolumn{6}{l}{group = None, q = -10}\\
base - pre & 0.02 & 0.03 & 1882.06 & 0.63 & 0.806 \\ 
  base - post & 0.03 & 0.03 & 1882.06 & 1.01 & 0.569 \\ 
  pre - post & 0.01 & 0.03 & 1880.01 & 0.40 & 0.916 \\ 
   \hline
\multicolumn{6}{l}{group = Isoch., q = -10}\\
base - pre & 0.65 & 0.03 & 1880.01 & 20.36 & $<$.001 \\ 
  base - post & 0.71 & 0.03 & 1880.01 & 22.11 & $<$.001 \\ 
  pre - post & 0.06 & 0.03 & 1880.01 & 1.76 & 0.185 \\ 
   \hline
\multicolumn{6}{l}{group = 1/f, q = -10}\\
base - pre & -0.20 & 0.03 & 1888.66 & -5.85 & $<$.001 \\ 
  base - post & -0.09 & 0.03 & 1888.66 & -2.80 & 0.0143 \\ 
  pre - post & 0.10 & 0.03 & 1880.01 & 2.97 & 0.009 \\ 
   \hline
\multicolumn{6}{l}{group = Rand., q = -10}\\
base - pre & 0.29 & 0.03 & 1882.06 & 9.75 & $<$.001 \\ 
  base - post & 0.33 & 0.03 & 1882.06 & 11.14 & $<$.001 \\ 
  pre - post & 0.04 & 0.03 & 1880.01 & 1.44 & 0.321 \\ 
   \hline
\multicolumn{6}{l}{group = None, q = 10}\\
base - pre & -0.18 & 0.03 & 1882.06 & -5.97 & $<$.001 \\ 
  base - post & -0.07 & 0.03 & 1882.06 & -2.53 & 0.031 \\ 
  pre - post & 0.10 & 0.03 & 1880.01 & 3.56 & 0.001 \\ 
   \hline
\multicolumn{6}{l}{group = Isoch., q = 10}\\
base - pre & 0.64 & 0.03 & 1880.01 & 19.96 & $<$.001 \\ 
  base - post & 0.53 & 0.03 & 1880.01 & 16.39 & $<$.001 \\ 
  pre - post & -0.11 & 0.03 & 1880.01 & -3.57 & 0.001 \\ 
   \hline
\multicolumn{6}{l}{group = 1/f, q = 10}\\
base - pre & 0.20 & 0.03 & 1888.66 & 5.85 & $<$.001 \\ 
  base - post & 0.15 & 0.03 & 1888.66 & 4.44 & $<$.001 \\ 
  pre - post & -0.05 & 0.03 & 1880.01 & -1.37 & 0.354 \\ 
   \hline
\multicolumn{6}{l}{group = Rand., q = 10}\\
base - pre & 0.41 & 0.03 & 1882.06 & 13.97 & $<$.001 \\ 
  base - post & 0.39 & 0.03 & 1882.06 & 13.27 & $<$.001 \\ 
  pre - post & -0.02 & 0.03 & 1880.01 & -0.72 & 0.752 \\ 
   \hline
\multicolumn{6}{l}{{\footnotesize Degrees-of-freedom method: kenward-roger}}\\

\multicolumn{6}{l}{{\footnotesize P value adjustment: tukey method for comparing a family of 3 estimates}}\\
\end{tabular}
\caption{\label{table:hq_table}Pairwise comparisons of $h(q)$ as a function of time and group at extreme values of q.}
\vspace{+0.1cm}
\end{table}

To understand the influence of various metronomes and a mechanical perturbation on gait variability, we investigated how the slope relating $q$ and $h(q)$ changed according to time and group. These results are presented in Table~\ref{table:slopes}. A general finding is that the $q \times h(q)$ slope differed between baseline and pre-perturbation for all conditions except the isochronous condition. In addition, the $q \times h(q)$ slope differed between baseline and post for the isochronous and 1/f conditions. Lastly, the pre-post differences in the $q \times h(q)$ slopes were only observed in the isochronous and 1/f conditions.

Collectively, these results emphasize a few key points related to the effects of various metronome types and fractal scaling in ISIs. First, the current results bolster previous observations that suggest isochronous and random metronomes produce gait patterns more consistent with pathology than with healthy gait \cite{kaipust2013gait,hunt2014influence,hausdorff1996fractal}. Second, the results while not entirely consistent, suggest that metronomes exert a profound effect over gait fractality, affecting the scaling properties of both very small and very large fluctuations, alike. The current slope analysis was limited to extreme values of $q$, but there is good reason to suspect that fluctuations of intermediate size would likewise be affected. Third, and surprisingly, our results did not produce a consistent pre-post perturbation effect with respect to either the central tendency of $h(q)$ or the $q \times h(q)$ slopes. There are several possible explanations for these pattern of results. Some conditions such as the random condition could have produced a floor effect, preventing successful reorganization of motor degrees of freedom to be detectable post-perturbation.    This is consistent with our hypothesis concerning 1/f noise where measurable pre-post perturbation effects were observed.  Another possibility is that post-perturbation incorporated in our analysis were too long. A limitation of MFDFA is time series need to quite long in order to obtain stable estimates. Hence, it is possible that the post-perturbation multifractality included too many strides after recovery to capture nuanced effects of perturbation. Future research should explore the use of a modified MFDFA procedure better suited for 'short' time series.

\begin{table}[ht]
\centering
\begin{tabular}{lrrrrl}
  \hline
contrast & estimate & SE & df & t & p \\ 
  \hline
\multicolumn{6}{l}{g = NONE}\\
base - pre & -0.01 & 0.00 & 1880.01 & -3.87 & 0.0003 \\ 
  base - post & -0.01 & 0.00 & 1880.01 & -2.08 & 0.0944 \\ 
  pre - post & 0.00 & 0.00 & 1880.01 & 1.84 & 0.1556 \\ 
   \hline
\multicolumn{6}{l}{g = Isoch.}\\
base - pre & -0.00 & 0.00 & 1880.01 & -0.23 & 0.9716 \\ 
  base - post & -0.01 & 0.00 & 1880.01 & -3.34 & 0.0024 \\ 
  pre - post & -0.01 & 0.00 & 1880.01 & -3.11 & 0.0053 \\ 
   \hline
\multicolumn{6}{l}{g = 1/f}\\
base - pre & 0.02 & 0.00 & 1880.01 & 6.90 & $<$.0001 \\ 
  base - post & 0.01 & 0.00 & 1880.01 & 4.27 & 0.0001 \\ 
  pre - post & -0.01 & 0.00 & 1880.01 & -2.54 & 0.0300 \\ 
   \hline
\multicolumn{6}{l}{g = Rand.}\\
base - pre & 0.01 & 0.00 & 1880.01 & 2.47 & 0.0359 \\ 
  base - post & 0.00 & 0.00 & 1880.01 & 1.25 & 0.4238 \\ 
  pre - post & -0.00 & 0.00 & 1880.01 & -1.26 & 0.4173 \\ 
   \hline
\multicolumn{6}{l}{{\footnotesize Degrees-of-freedom method: kenward-roger}}\\

\multicolumn{6}{l}{{\footnotesize P value adjustment: tukey method for comparing a family of 3 estimates}}\\
\end{tabular}
\caption{\label{table:slopes}Analysis of $q\times h(q)$ slopes as a function of time and group.}
\vspace{+0.1cm}
\end{table}

\section{Next-Generation Networked Metronomes}
\label{sec:nextgen}

In this section, we describe directions for building the next generation of metronomes, featuring advanced networking capabilities in conjunction with technologies, such as Augmented Reality (AR), to provide an immersive and interactive quality of experience to participants~\cite{shannigrahi2020next}. 

\subsection{Smart Phone Metronome Applications}

A direction for the implementation of gait metronomes nowadays is through smart phone applications. Such applications will be installed on the research participants' phones, while parameters, such as the type and frequency of the auditory beat, can be adjusted as input parameters of the applications. Such metronomes enable flexible mobility opportunities, so that experiments can be performed outside of laboratory spaces in daily living contexts (e.g., walking on a university campus, working out at a park). This is particularly important because laboratories may restrict the flexibility of the performed experiments (e.g., treadmills constrain participants to walk on fixed directions). On the other hand, daily living contexts can offer insights to habitual behaviors, which may not be feasible in laboratory spaces.

\subsection{Augmented Reality and ``Gamified'' Metronomes}

\begin{figure}
    \centering
    \includegraphics[width=1\linewidth]{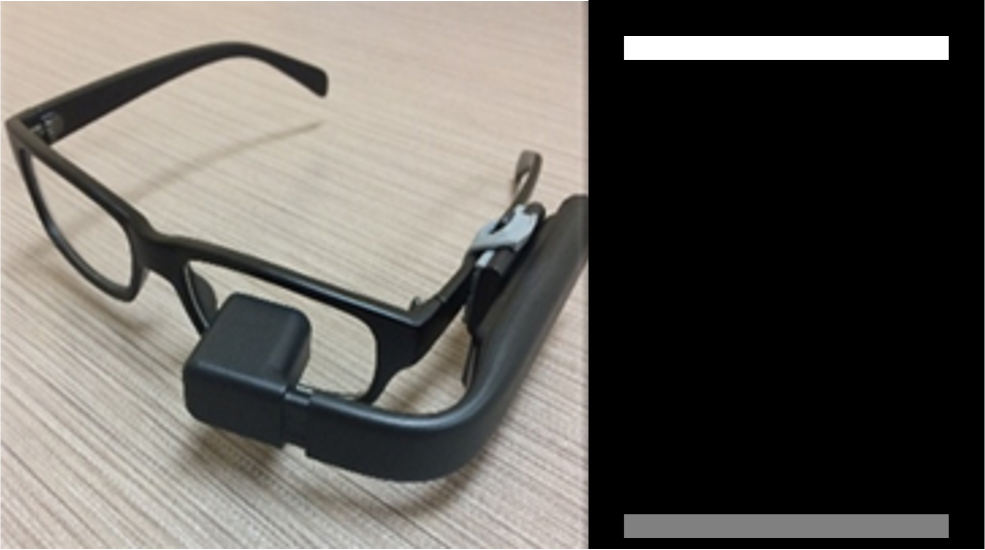}
    \caption{Prototype of visual metronome. A moving bar depicted on right hand side is displayed on a video screen mounted to a non-prescription pair of glasses.}
    \label{fig:glasses} 
    \vspace{+0.5cm}
\end{figure}

So far, we have focused on auditory gait metronomes. The auditory stimuli is discreet and limited in nature. To this end, gait metronomes can be augmented with visual effects, so that they provide visual (continuous) stimuli to research participants, offering new opportunities for gait-related research. In Figure~\ref{fig:glasses}, we present a prototype of a visual metronome. This metronome is implemented through a pair of glasses that participants wear. These glasses are powered and are directly connected to a tablet device. The stimuli are visualized through a bar that moves vertically. When the bar reaches its highest point, the participants are instructed to touch the ground with their feet as they are taking a step. 

\begin{figure}
    \centering
    \includegraphics[width=1\linewidth]{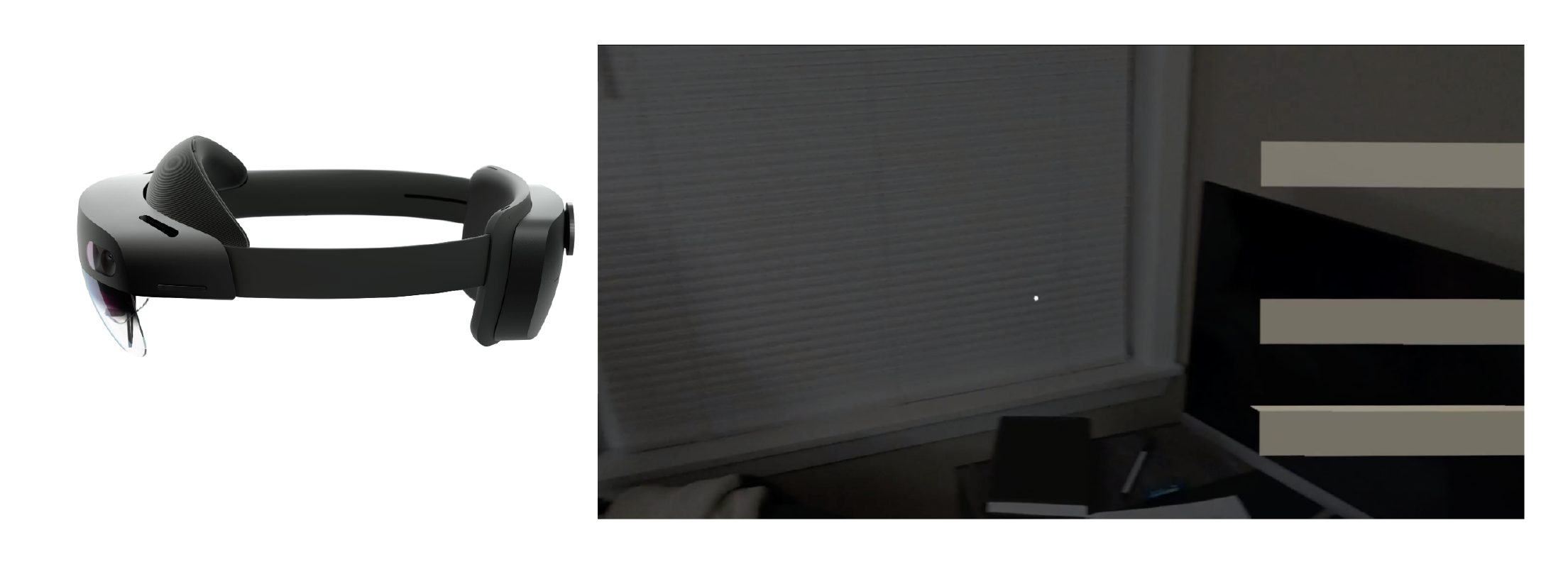}
    \caption{Prototype of a AR-capable metronome developed based on Unity and deployed on a Hololens 2 headset.}
    \label{fig:AR} 
    \vspace{+0.5cm}
\end{figure}

This visual metronome prototype, however, requires that participants carry a tablet device with them while walking. Technologies to connect the glasses wirelessly (e.g., Bluetooth, WiFi) to the tablet may be useful, however, still such glasses have limited capabilities in terms of graphics. A direction that is worthy of exploring is the integration of AR to provide immersive graphical experience that enhances the research participant's perception of the world. In Figure~\ref{fig:AR}, we present an early prototype of an AR-based metronome that we implemented based on Unity. We have deployed this prototype on a Hololens 2 headset. This metronome is a direct conversion of the metronome of Figure~\ref{fig:AR} to an AR environment.

Leveraging AR, a direction to explore is ``gamifying'' gait metronomes in order to provide an interactive, gaming-like experience to research participants. For example, participants may be instructed to synchronize their steps with visual targets (e.g., collide with 3D bulls-eye objects). Another example may be a metronome, where participants need to step on 3D tiles of a certain color or shape that the AR headset projects on the floor. At the same time, participants can receive real-time feedback about their performance in this gaming experience (e.g., through messages displayed on the headset's screen), being: (i) encouraged to step on 3D bulls-eye objects or certain 3D tiles if they do not follow the instructions of the game; and (ii) praised for their performance if they follow the instructions of the game.

\subsection{Cloud and Edge Computing Technologies}

Gait metronomes coupled with wireless capabilities and potentially based on AR may need to utilize remote computing resources for computationally expensive operations (e.g., 3D rendering, object detection). Cloud computing technologies could facilitate the execution of computationally expensive operations on remote computing resources. However, cloud computing resources may result in high response times, which can be problematic in cases of low-latency applications, such as interactive AR. To this end, edge computing~\cite{shi2016edge} has emerged as a paradigm that features computing resources physically close to users (at the edge of the network), subsequently resulting in fast response times.

Edge computing solutions open opportunities for future computer networking research. Such opportunities include mechanisms for the allocation and discovery of computing resources at the edge~\cite{mastorakis2019towards, mastorakis2020dlwiot}, as well as the dynamic offloading of computational tasks at the edge~\cite{mastorakis2020icedge}. For example, if multiple edge servers are available, we may need consecutive computational tasks to be offloaded to the same edge server. This is due to the fact that the processing of these tasks may determine whether a participant follows the provided instructions (e.g., whether the participant is stepping on projected 3D tiles of the right color) not only momentarily, but over a certain period of time. The processing outcome of such tasks will subsequently determine the feedback to be provided to the participant (e.g., the message to be displayed on the headset's screen). In addition, mechanisms for in-network computing coupled with advanced wireless capabilities (e.g., LTE/5G) could further increase the available bandwidth and reduce response times~\cite{nour2020compute, ullah2020icn}. 

\section{Conclusion and Future Work}
\label{sec:conclusion}

In this paper, we have presented a preliminary study investigating the utility of different metronomes as assistive devices. Our results showed that, when followed by healthy adults, metronomes typically used for rehabilitation (i.e., isochronous metronomes) produce gait patterns more consistent with pathological gait than variability observed in typical walking. In contrast, metronomes based on healthy gait patterns left those patterns unperturbed, even after experiencing a strong mechanical perturbation. Lastly, we provided initial plans for enhancing metronomes as assistive devices through the use of advanced networking and computing technologies as well as augmented reality. Our vision is that these technologies will advance these promising tools by allowing better graphics, faster calculations, and more freedom of use. Our goal is to increase the efficacy of metronomic rehabilitation tools by: (1) using metronomes that take into consideration the natural regularities in healthy gait, (2) providing a better user experience in a broader range of settings. Ultimately, these improvements may encourage broader use of these tools and prevent falls.

\section*{Acknowledgements}

This work was supported by the National Institutes of Health (NIGMS/P20GM109090), the National Science Foundation (CNS2016714), and the University of Nebraska Collaboration Initiative.

\bibliographystyle{IEEEtran}  
\bibliography{bib}



\end{document}